\documentclass[aps,prb,reprint,superscriptaddress]{revtex4-1}
\pdfoutput=1
\usepackage{graphicx}
\usepackage{bm}
\usepackage{amsmath}
\usepackage{comment}

\begin{document}


\title{Charging dynamics of single InAs quantum dots under both resonant and above-band excitation}


\author{Gary R. Lander}
\email[]{glander@mix.wvu.edu}
\affiliation{Department of Physics and Astronomy, West Virginia University, Morgantown, WV 26506, USA}

\author{Samantha D. Isaac}
\affiliation{Department of Physics and Astronomy, West Virginia University, Morgantown, WV 26506, USA}

\author{Disheng Chen}
\affiliation{Department of Physics and Astronomy, West Virginia University, Morgantown, WV 26506, USA}

\author{Samet Demircan}
\affiliation{Department of Physics and Astronomy, West Virginia University, Morgantown, WV 26506, USA}


\author{Glenn S. Solomon}
\affiliation{Joint Quantum Institute, National Institute of Standards and Technology, \& University of Maryland, Gaithersburg, MD, USA.}

\author{Edward B. Flagg}
\email[]{edward.flagg@mail.wvu.edu}
\affiliation{Department of Physics and Astronomy, West Virginia University, Morgantown, WV 26506, USA}


\date{\today}

\begin{abstract}
We investigate the charging dynamics in epitaxially grown InAs quantum dots under resonant excitation with and without additional low-power above-band excitation. Time-resolved resonance fluorescence from a charged exciton (trion) transition is recorded as the above-band excitation is modulated on and off. The fluorescence intensity varies as the QD changes from charged to neutral and back due to the influence of the above-band excitation. We fit the transients of the time-resolved resonance fluorescence with models that represent the charging and neutralization processes. The time dependence of the transients indicate that Auger recombination of resonantly excited trions is largely responsible for neutralization of the charged state when the above-band excitation is off. The addition of above-band excitation revives the resonance fluorescence signal from the trion transition.
We conclude that the above-band laser excites charges that relax into the bound state of the quantum dot via two different charge transport processes. The captured charges return the QD to its initial charge state and allow resonant excitation of the trion transition. The time dependence of one charge transport process is consistent with ballistic transport of charge carriers excited non-local to the QD via above-band excitation. We attribute the second charge transport process to carrier migration through a stochastic collection of weakly-binding sites, resulting in sub-diffusion-like dynamics.
\end{abstract}

\pacs{need to find these}

\maketitle


\section{introduction}
Sources of single, indistinguishable photons are a promising candidate for implementation of quantum information protocols \cite{zeilinger_three-particle_1997, pan_experimental_1998, knill_scheme_2001, fattal_entanglement_2004, moehring_entanglement_2007}. Semiconductor quantum dots (QDs) can act as sources of single photons to be utilized in these protocols, but certain experimental factors can complicate their single photon emission. Spectral diffusion broadens the emission line shape \cite{robinson_light-induced_2000}, and in samples without independent electrical control of the QD charge, blinking occurs when the QD changes charge state \cite{chen_characterization_2016, nguyen_photoneutralization_2013, davanco_multiple_2014, hu_defect-induced_2015}. Even with electrical control of the QD charge, Auger recombination from the trion state may neutralize the dot \cite{kurzmann_auger_2016}. Resonant excitation of either a neutral or charged quantum dot can cause a transition to the opposite charge state, which greatly diminishes the time-averaged fluorescence and reduces a dot's suitability to act as an efficient photon source \cite{metcalfe_resolved_2010, nguyen_photoneutralization_2013, jons_two-photon_2017}. A counter to this effect is the application of a low-power above-band-gap laser that supplies the local charge environment with extra charge carriers \cite{gazzano_effects_2018, nguyen_photoneutralization_2013}. These charge carriers can be captured by either a charged QD, resulting in neutralization and allowing resonant excitation of the exciton state, or by a neutral QD, allowing resonant excitation of the trion state. The exact processes by which the above-band excited carriers arrive in the QD is so far uncertain. Understanding those processes will inform the design of future QD-based sources of single photons.

Here we investigate the charge dynamics of a quantum dot under both resonant and above-band excitation. We measure the rise and fall of the fluorescence of a resonantly excited trion transition as the above-band laser is turned on and off. We characterize the time-dependent dynamics as a function of the two excitation powers using two models that describe the time evolution of the fluorescence after the above-band laser either turns on or turns off. We conclude that Auger recombination is the primary mechanism by which the quantum dot becomes neutralized, by which we mean it changes from a charged state (e.g. the trion or a single bound charge) to a neutral state (e.g. the exciton, biexciton, or the empty QD). After such a neutralization event, the resonance fluorescence is absent and the QD remains neutral until it captures another charge. The time dependence of the fluorescence rise and decay indicate that there are at least two processes that provide charge carriers to be captured by the QD. One is ballistic super-diffusion of hot carriers generated near the surface of the sample by the above-band laser. The other is sub-diffusion of carriers through a stochastic collection of weakly-binding transport sites which are likely an ensemble of shallow states in the semiconductor.

%
\section{experimental setup}
\begin{figure}[t]
	\includegraphics[width=3.4in]{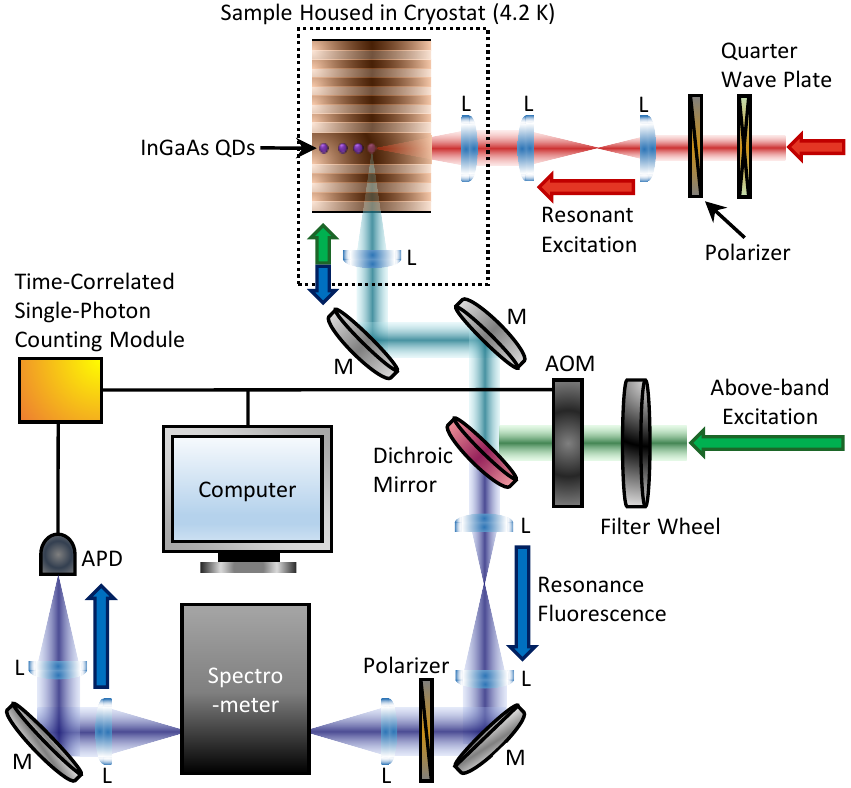}
	\caption{\label{fig:experimental setup}  A schematic of the experimental setup. The sample is housed in a cryostat kept at approximately 4.2 K. The red path depicts resonant excitation, the green above-band excitation, the blue resonance fluorescence, and the cyan where the above-band excitation and resonance fluorescence overlap in the confocal optical path. M-mirror, L-lens, AOM-acousto-optical modulator, APD-avalanche photodiode.}
\end{figure}

Our sample consists of epitaxially grown self-assembled InAs quantum dots embedded in the GaAs spacer of a planar microcavity defined by two distributed Bragg reflectors (DBRs). The resonant excitation laser is focused into the waveguide mode of the microcavity from the side and the fluorescence is collected normal to the sample surface, perpendicular to the resonant excitation direction, which minimizes collection of laser scattering \cite{muller_resonance_2007-1, flagg_resonantly_2009, chen_resonance_2017}. The above-band excitation is provided by a helium-neon laser with a wavelength of 632 nm that is focused onto the QD normal to the sample surface and confocal with the resonance fluorescence collection path. The above-band excitation power used throughout is too low to cause any detectable fluorescence on its own. Only when the resonant and above-band lasers are both on does the QD emit fluorescence. The above-band excitation is modulated with an acousto-optic modulator (AOM) while the resonant laser intensity remains constant. The above-band light is filtered from the fluorescence via a dichroic mirror, long-pass filters, and a spectrometer. Ultimately, the fluorescence is incident on a silicon-based avalanche photodiode capable of detecting single photons with a time resolution of approximately 500 ps. The photon arrival times are then recorded via a time-correlated single-photon counting module (TCSPC). The voltage signal used to modulate the AOM is also used to create a trigger sent to the TCSPC. The time-resolved fluorescence can then be constructed as a histogram of the photon arrival times relative to the preceding modulation trigger. A diagram of the experimental setup is shown in Fig.~\ref{fig:experimental setup}.
\section{data}
\begin{figure*}[!t]
	\includegraphics[width=7in]{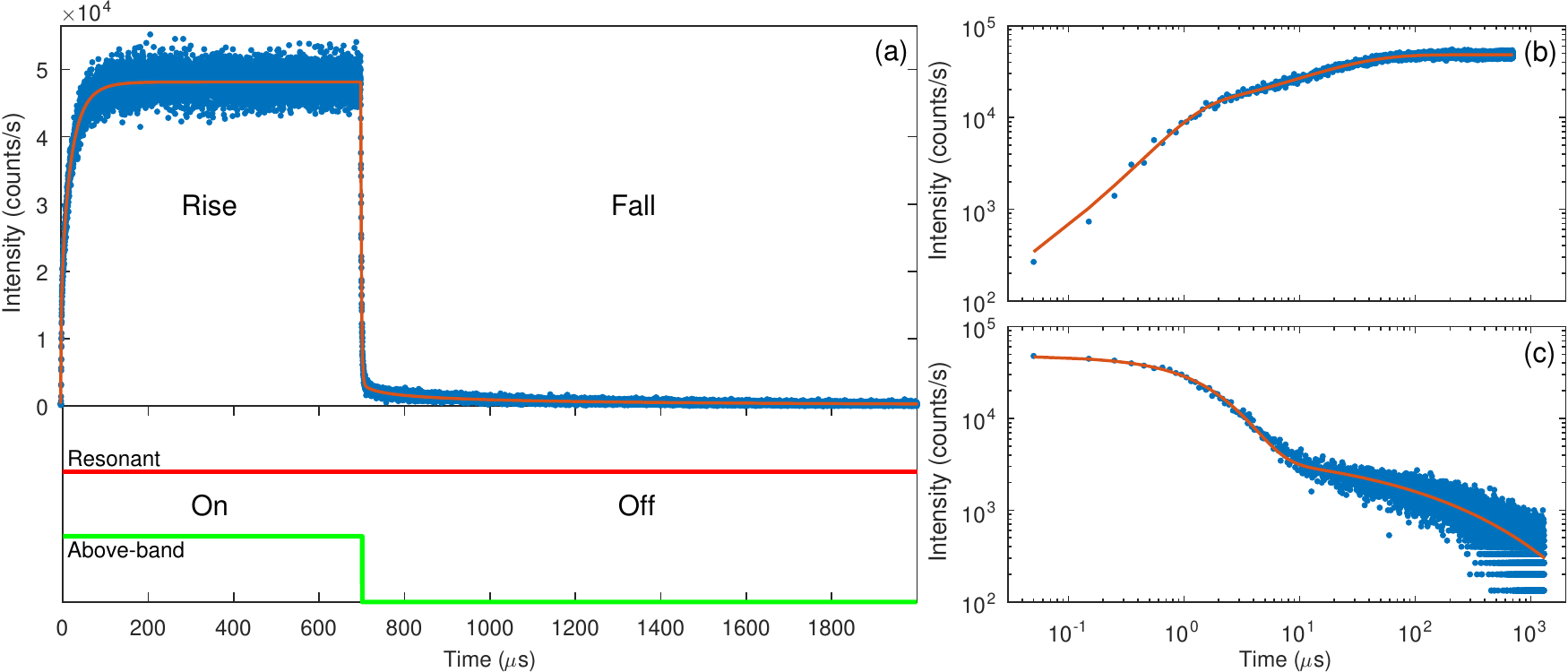}
	\caption{\label{fig:data}  (a) Typical time-resolved resonance fluorescence. For the displayed data the resonant excitation power is 2.5 $\mu$W, while the above-band power is 0.27 $\mu$W. The orange line is the fit to the data. Below, the green line indicates when the above-band excitation power is on and off. The red line indicates the resonant excitation power, which is always on. (b) A fit to the rise section on a log-log scale. (c) A fit to the fall section on a log-log scale.}
\end{figure*}

Without the above-band excitation, the fluorescence signal is negligibly small despite constant resonant excitation. When the above-band laser is turned on, the fluorescence signal increases asymptotically to a steady state value.  When the above-band laser is turned off, the fluorescence signal decreases asymptotically to zero. Figure~\ref{fig:data}(a) shows typical time-resolved resonance fluorescence data for a given power of both above-band and resonant excitation. The time scale for the system to reach steady state is on the order of hundreds of microseconds for the rise transient after the above-band excitation is turned on, and tens to hundreds of microseconds for the fall transient after the above-band excitation is turned off.  Time-resolved fluorescence was recorded for multiple different powers of both the resonant and above-band lasers. In all cases the above-band power was too weak to cause fluorescence on its own. The laser powers spanned approximately two orders of magnitude, defining a two-dimensional acquisition space.  The rise and fall sections of the example data in Fig.~\ref{fig:data}(a) are shown separately in Figs.~\ref{fig:data}(b-c), respectively, using a log-log scale to illustrate the short-time dynamics. The models discussed below fit the data well at all time scales and particular model curves are shown on top of the data in Figs.~\ref{fig:data} (a-c).
\section{discussion}
The models we use involve the shaped exponential function, $e^{-(\alpha t)^{\beta}}$, which fits well a large variety of relaxation phenomena in complex condensed-matter systems \cite{williams_non-symmetrical_1970, bohmer_nonexponential_1993, phillips_stretched_1996, klafter_relationship_1986, anderssen_kohlrausch_2004, bodunov_origin_2017, doi:10.1002/9783527622979.ch11}. When $0<\beta<1$, the above mathematical form is referred to as a stretched exponential, whereas when $1<\beta<2$ it's referred to as a compressed exponential. The time-resolved fluorescence intensity is fit with either the sum of a normal and a stretched exponential (for the fall), or the numerical solution of a differential equation containing capture rates that saturate with stretched and compressed exponential time dependencies (for the rise). The details of the models are discussed below. We extract the model parameters and discuss their dependencies on laser power. Throughout this letter we will refer to $\alpha$ as a scale parameter and $\beta$ as a shape parameter.

The stretched exponential $(0<\beta<1)$  has been sporadically used for over 160 years to explain relaxation phenomenon in condensed matter systems. The first known instance was carried out by the physicist Rudolf Kohlrausch to explain relaxation of residual charge from a glass Leiden jar \cite{kohlrausch_r._theorie_1854}. The shaped exponential has mostly been used as a phenomenological fit and there has been much difficulty in applying direct physical connections to the scale and shape parameters $\alpha$ and $\beta$. Despite the difficulty in making clear physical correlations to the parameters, there has been some headway for the development of a physical model that explains the shaped exponential behavior in the solid state. Klafter \textit{et al.} illustrated three different physical approaches that result in stretched exponential charge transfer in the solid state \cite{klafter_relationship_1986}. The similarity between the models presented by Klafter \textit{et al.} shows that either multiple different pathways into a trap or multiple different pathways out of a trap are required. Random numbers and physical shapes of the pathways for charges to migrate through, due to being part of a stochastic environment, seem paramount for the emergence of stretched exponential relaxation. Migration through such a stochastic environment results in slower than exponential relaxation via this channel.

In 2003, Sturman \textit{et al.} described how stretched exponential relaxation in the solid state can result from charge carriers migrating through an environment of stochastically distributed transport sites before ultimately relaxing into a trap \cite{sturman_origin_2003}. For clarity we briefly summarize their results here. Consider an environment containing a stochastic distribution of many weakly-binding potential wells called transport sites plus a few strongly-binding potential wells called traps. From an initial transport site, a charge carrier will naturally have a higher probability to hop to closer transport sites as opposed to farther ones, defining random pathways through which charge carriers will likely migrate. Some of these pathways lead to transport sites in the close vicinity of traps, resulting in a significant probability of the charge being captured by the trap. Simulations of such a system show that the process of slow migration through such a stochastic environment results in stretched exponential relaxation into the traps; see Ref.~[26] for more details. A stretched exponential can be mathematically represented as a linear sum of normal exponentials with a certain weighting function \cite{johnston_stretched_2006}. Due to different relaxation rates associated with different paths through the transport sites, the relaxation is described by a sum of many exponential decays, and more compactly a single stretched exponential term to describe the net process.

The compressed exponential $(1<\beta<2)$ has been less widely used to describe charge relaxation in the solid state. However, Ref.~[24] gives examples of instances where compressed exponential relaxation in the sold state is observed. The common theme is the presence of an external driving force resulting in faster than exponential relaxation. Morishita describes compressed exponential relaxation dynamics in liquid silicon above 1200 K \cite{morishita_compressed_2012}. He attributes the compressed exponential relaxation to ballistic-like motion of high-energy carriers, similar to the material described by Bouchaud. This type of motion can be described as super-diffusion, due to the associated faster-than-diffusion-like behavior.

In the analysis below we will describe how the shaped exponential growth and decay behavior of the time-resolved fluorescence is consistent with either ballistic transport of carriers or migration through a stochastic distribution of transport sites. We start by discussing the fall section of the fluorescence, depicted in Fig. 2(c), because without the above-band laser the behavior is less complicated.
\subsection{Fluorescence Fall}
At the beginning of the fall section depicted in Fig.~\ref{fig:data}(c) the above-band excitation is turned off and the fluorescence signal decreases asymptotically to zero as the quantum dot becomes neutralized. There is an initial fast decay in the resonance fluorescence followed by a slower decay at long times. The fluorescence intensity data were fit with the sum of a normal exponential and a stretched exponential. The functional form is:
\begin{equation}
I=A_1e^{-\alpha_1t}+A_2e^{-(\alpha_2t)^{\beta_2}} \label{eqn:fall} \\                                                               
\end{equation}
$A_1$ and $A_2$ are amplitudes where $A_1+A_2$ is the steady-state value of the intensity when the above-band excitation is on, $\alpha_1$ is a neutralization rate while $\alpha_2$ is a scale parameter, and the exponent $\beta_2$ is a shape parameter.

Figure~\ref{fig:data}(c) shows an example of this model fitting the fall data for a certain pair of laser powers. The fast decay corresponds to the normal exponential term in the fit, while the slower decay corresponds to the stretched exponential term. As described below, the power dependence of the fast decay indicates that it results from Auger recombination of the excited trion state. We attribute the slower decay to recharging of the dot by capture of charge carriers from a reservoir whose own population is decaying with a stretched exponential dependence.
\begin{figure}[t]
	\includegraphics[width=3.4in]{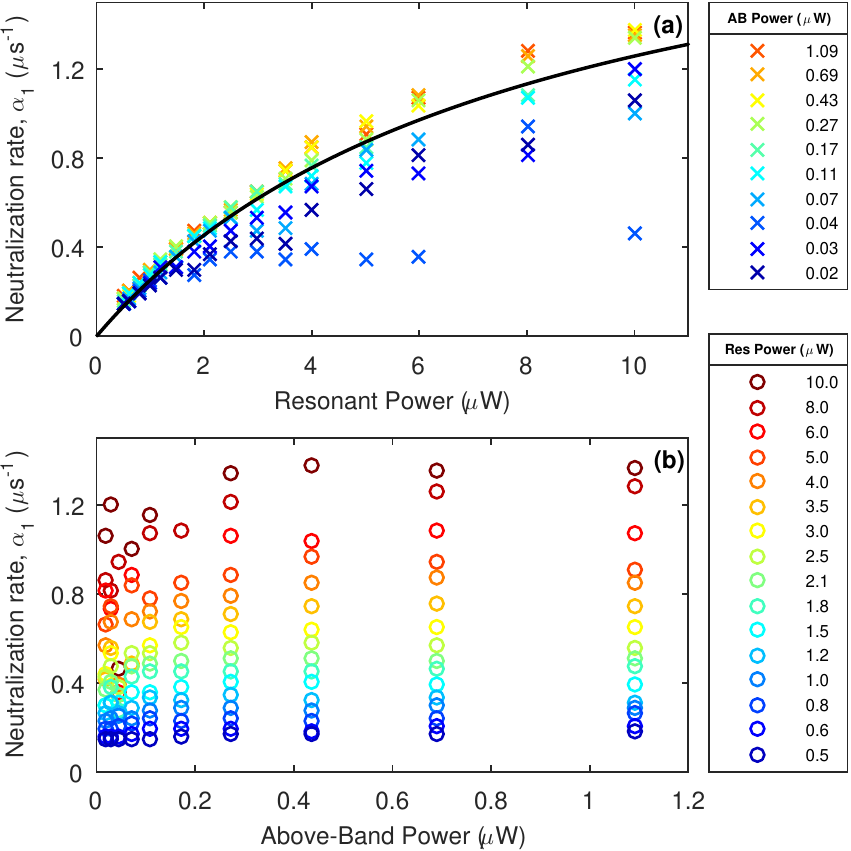}
	\caption{\label{fig:alpha1} Fast neutralization rate plotted vs.~ (a) resonant laser power and (b) above-band laser power. The legends and color specify the power of the excitation laser not represented by the horizontal axis. The saturation curve shown in (a) is the best fit to the entire data set shown. A saturation power of 7.9 $\pm$ 1.4 $\mu$W and an Auger recombination rate of 2.3 $\pm$ 0.2 $\mu$s$^{-1}$ was extracted from the fit.}
\end{figure}

Figure~\ref{fig:alpha1}(a) plots the neutralization rate $\alpha_1$ versus resonant laser power, and it increases sub-linearly with increasing resonant power. Its power dependence is similar to that of the excited state population of a resonantly excited two-level system, which saturates as the laser power increases. The two-level system in this case is the single-charge/trion system of the charged QD. Since $\alpha_1$ has a similar saturation behavior, it implies that the rate of the neutralization process represented by the normal exponential is proportional to the trion population in the QD. Such a dependence is consistent with Auger recombination of the trion, which ejects the extra charge carrier and neutralizes the QD. Thus, Auger recombination is responsible for the fast, normal exponential decay. This conclusion is consistent with other experiments that have measured time-dependent charge state dynamics \cite{gazzano_effects_2018, kurzmann_auger_2016}. The values of $\alpha_1$ in Fig.~\ref{fig:alpha1}(a) are fit with a saturation curve, $\alpha_1 = (\Gamma_A P) / 2 (P+P_\mathrm{sat})$, where $\Gamma_A$ is the Auger recombination rate for the trion, $P$ is the resonant laser power, and $P_{\mathrm{sat}}$ is the saturation power. We obtained a saturation power of 7.9 $\pm$ 1.4 $\mu$W and an Auger recombination rate of 2.3 $\pm$ 0.2 $\mu$s$^{-1}$, which is in very close agreement to the value of 2.3 $\mu$s$^{-1}$ found by Kurzmann \textit{et al.} \cite{kurzmann_auger_2016} in InAs QDs. We observe little dependence of the Auger recombination rate as a function of above-band power as seen in Fig.~\ref{fig:alpha1}(b).

If Auger recombination was the only process involved in the fluorescence decay, then only one term would be needed in the model. The presence of the additional slow decay implies an additional \textit{recharging} process that weakly counteracts the Auger neutralization of the QD. The recharging process itself must decay with time or the fluorescence would reach a non-zero steady-state value. The stretched exponential time dependence of the slow decay is consistent with the expected decay of charges migrating through a stochastic distribution of transport sites into the dot \cite{sturman_origin_2003}, as discussed above. In this case, the transport sites are likely shallow impurities, iso-electronic dopants, or other defects in the GaAs host, which can weakly bind charge carriers at the low temperature of the cryostat. The transport sites close to the QD comprise a reservoir that can recharge the QD under investigation as long as they contain charges. While the above-band excitation is on, the reservoir is continually repopulated. When the above-band excitation turns off, the charge population of the reservoir begins to decay. As the population decays, so too will the average value of the QD charge state. Thus the reservoir decay is reflected in the long-time decrease of the measured fluorescence.

\begin{figure}[t]
	\includegraphics[width=3.4in]{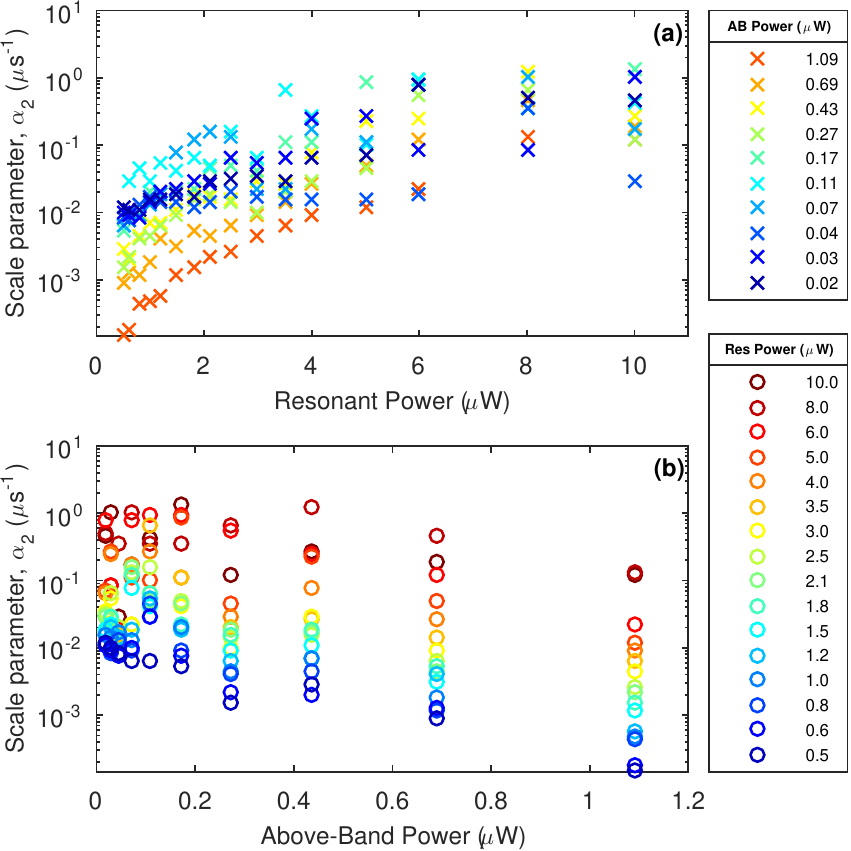}
	\caption{\label{fig:alpha2} Scale parameter $\alpha_2$ vs.~ (a) resonant laser power and (b) above-band laser power. The legends and color specify the power of the excitation laser not represented by the horizontal axis.}
\end{figure}

The scale parameter $\alpha_2$ increases with resonant laser power regardless of the above-band laser power, as seen in Fig.~\ref{fig:alpha2}(a). The resonant laser can excite carriers weakly bound to transport sites to higher energy continuum states in the semiconductor host, which increases the rate of reservoir depletion after the above-band excitation is turned off, and hence the value of $\alpha_2$ increases with resonant laser power. Figure~\ref{fig:alpha2}(b) shows that $\alpha_2$ decreases as a function of above-band power. Each transport site that is a component of the sub-diffusion-like reservoir can only hold (on average) one charge. When many transport sites are filled with higher above-band power, charges move to adjacent transport sites with an average slower rate due to the system being 'clogged' with charges occupying transports sites on the way to the QD. Thus, when the above-band power is high before it is shut off, we see a slower depletion of the reservoir via the time-resolved resonance fluorescence measured from the QD.
\subsection{Fluorescence Rise}
At the beginning of the rise section depicted in Fig.~\ref{fig:data}(b), the quantum dot starts out neutral and resonant excitation of the trion transition does not cause any fluorescence. When the above-band laser is turned on, charges are excited and begin to relax into the bound state of the quantum dot. The time-averaged charge state of the quantum dot changes from neutral to charged. Towards the end of the rise section, the time-averaged charge occupation in the quantum dot has reached a steady state value and the fluorescence intensity remains at maximum brightness as long as the above-band laser is on. However, the full transient of the rise spans tens to hundreds of microseconds, which implies a situation where the charges excited by the above-band laser do not immediately become bound to the quantum dot. Since excitation of charges occurs within a few ps of the above-band laser turning on, it follows that it is the capture process that is delayed rather than excitation. The reason charge capture is delayed is that the charge carriers are excited nonlocal to the QD layer.

The above-band laser (632 nm) is focused normal to the sample surface with a beam waist of approximately 4.5 $\mu$m. The top DBR is comprised of alternating layers of AlGaAs and GaAs. The penetration depth, the depth at which the laser intensity decays to $1/e$ of its surface value, is approximately 340 nm for the above-band photons entering the heterostructure \cite{rakic_modeling_1996}. However, the QD layer exists slightly less than 2300 nm below the sample surface. A detailed to-scale schematic illustrating the penetration depth of the above-band laser in the sample can be found in the Supplemental Material. The generated charges can reach the monitored QD via several different transport mechanisms. Depending on the nature of the mechanisms involved, the charges may migrate with sub-diffusion, super-diffusion, or regular diffusion-like dynamics.

As discussed in the previous section, we suggest there exists a stochastic distribution of transport sites through which charge carriers can migrate to relax into the QD . The effective reservoir supplying the QD with charge is the collection of transport sites adjacent to, or in the near vicinity of the QD such that there is a substantial probability for charges occupying those transport sites to become bound to the QD within the time span of the observed transients. These transport sites are part of a larger network of transport sites through which charges migrate with a sub-diffusion-like behavior. Thus, we anticipate the existence of a charging process that has stretched exponential time dependence.

As shown in Refs.~[24,28], the ballistic transport of hot carriers results in super-diffusion which in turn is mathematically represented by compressed exponential time dependence. Almost 90 percent of the above-band induced hot charge carriers are created within 700 nm of the sample surface. The photon energy of the above-band laser translates to an initial excited carrier velocity of 800 nm/ps. The QD layer lies approximately 2300 nm below the sample surface. Hot carriers in bulk GaAs have an average lifetime of approximately 4 or 5 ps \cite{xu_picosecond_1984}. In that time a hot carrier can travel a few thousand nanometers. Thus, it is feasible that hot carriers excited in the first 700 nm from the sample surface can reach the QD layer before relaxing to the gamma point, and ultimately becoming bound to the QD.  We hypothesize the existence of a super-diffusion-like pathway into the dot that is supplied by hot carriers created non-local to the QD layer by the focused above-band laser and its associated shallow penetration depth. Some carriers have ballistic trajectories that set them near the QD layer as they cool to the gamma point. This local volume of the bulk semiconductor host thus acts as an additional effective reservoir to supply the dot with charge, and its population saturates with compressed exponential time dependence due to the super-diffusion-like nature by which it is supplied with charge carriers. Thus, we anticipate a charging process that has compressed exponential time dependence.
\begin{figure*}[t]
	\includegraphics[width=7in]{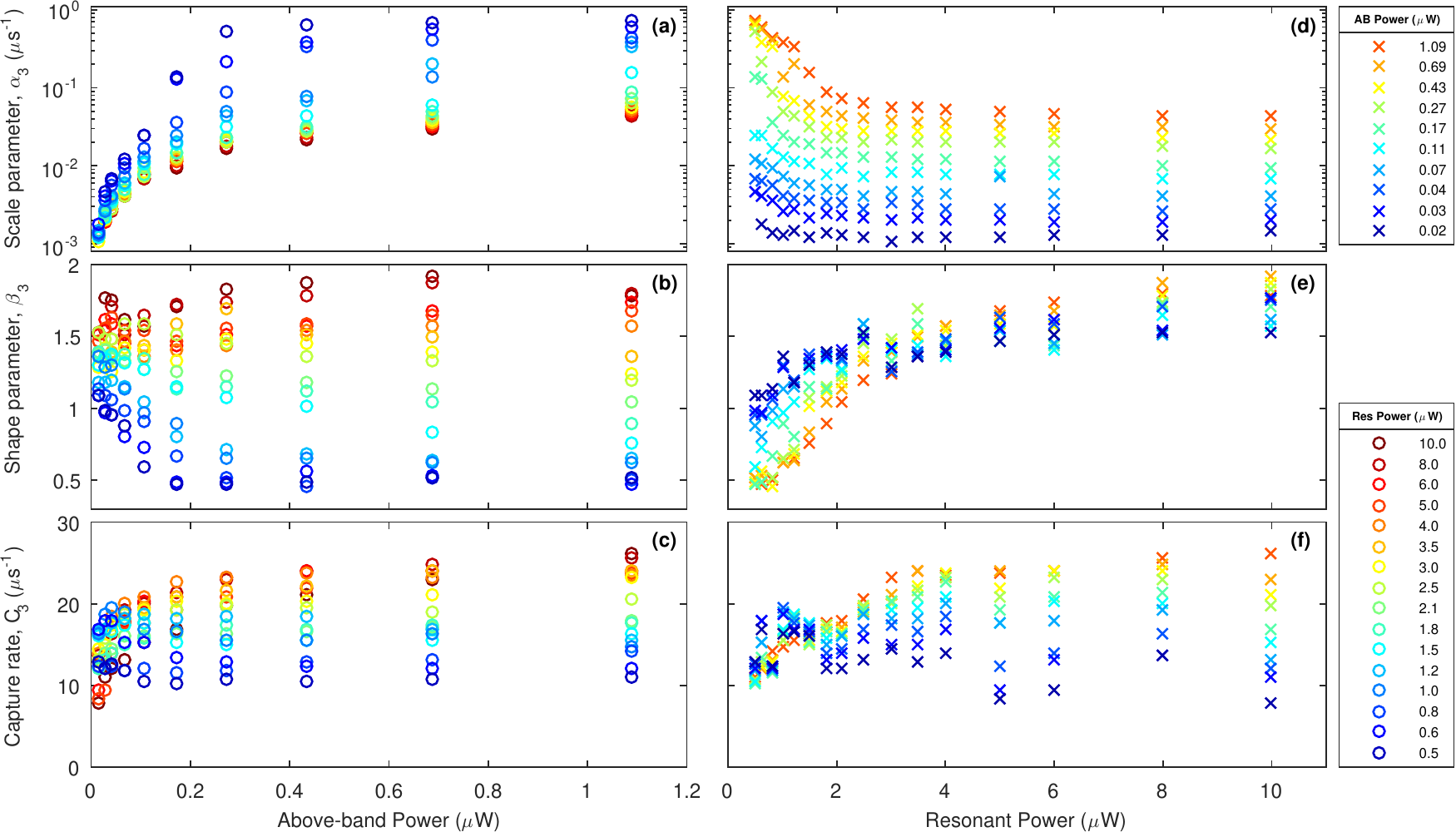}
	\caption{\label{fig:superdiffusion} Model parameters associated with the super-diffusion process as a function of (a-c) above-band power and (d-f) resonant laser power. The legends and color specify the power of the excitation laser not represented by the horizontal axis.}
\end{figure*}

We numerically model the charging dynamics of the rise section of the data using the solution to a differential equation for the time derivative of the ensemble-averaged QD charge population, $p$. We include the two charging processes discussed above as terms with stretched and compressed time dependence that are proportional to the unoccupied charge state in the QD, $(1-p)$. Finally, we also include a neutralization term that is proportional to $p$. The resulting differential equation is:
\begin{align}
&\frac{d p}{{dt}}=-\Gamma_N p \nonumber \\
&+ [C_3(1-e^{-(\alpha_3t)^{\beta_3}})+C_4(1-e^{-(\alpha_4t)^{\beta_4}})](1-p) \label{eqn:rise}
\end{align}

The ensemble-averaged charge population in the dot, $p$, can take on a value between 0 (neutral) and 1 (charged). $\Gamma_N$ is the neutralization rate per unit charge in the dot, $C_3$ and $C_4$ are long time capture rates per unoccupied charge $(1-p)$ in the dot for two separate charging pathways into the dot, $\alpha_3$ and $\alpha_4$ are scale parameters associated with the two separate pathways, and $\beta_3$ and $\beta_4$ are shape parameters. During development of the model, we assumed a neutralization rate proportional to the ensemble-averaged charge population in the dot and one or two capture rates that were either constant in time or had saturating time dependencies. We ultimately obtained the best fits using two charging terms: one that fills with stretched exponential time dependence and one with compressed exponential time dependence. The rate at which charges enter the dot should be proportional to not only the unoccupied charge in the dot, but also the population of the effective reservoir supplying the associated charging pathway with charges. $\alpha_3$ and $\alpha_4$ represent the effective rates of change of charge population in the two reservoirs after the above-band excitation turns on. During the fitting process, $\beta_3$ and $\beta_4$ always obtained values between 0 and 2. The term with index 3 is associated with the super-diffusion-like process, while the term with the index 4 is associated with the sub-diffusion-like process.
\begin{figure*}[t]
	\includegraphics[width=7in]{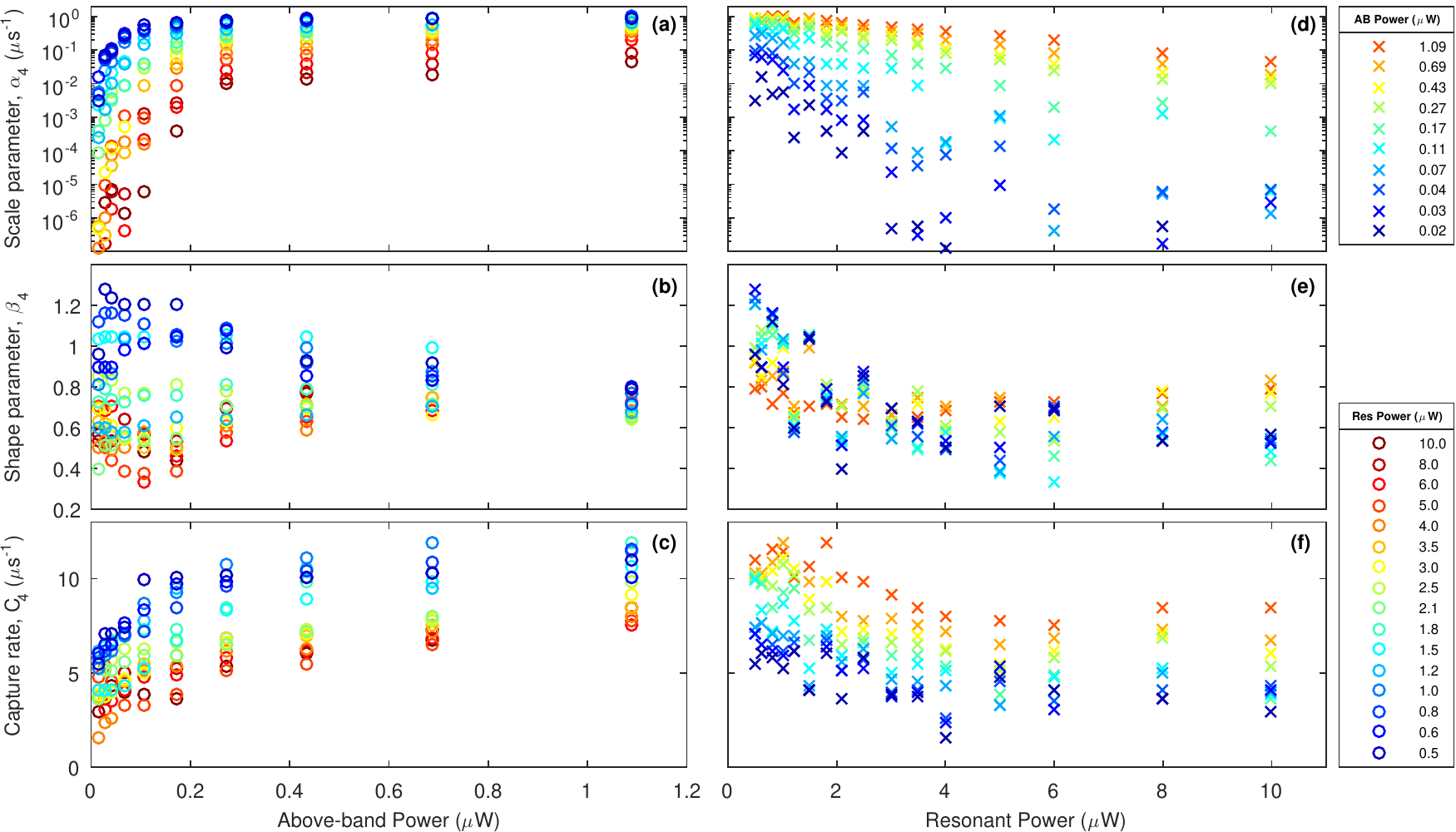}
	\caption{\label{fig:subdiffusion} Model parameters associated with the sub-diffusion process as a function of (a-c) above-band power and (d-f) resonant laser power. The legends and color specify the power of the excitation laser not represented by the horizontal axis.}
\end{figure*} 

We should expect that as the above-band laser power increases, the model parameters will change to hasten the rise of the fluorescence. The resonant laser causes both fluorescence and Auger recombination. It also suffuses the entire waveguide region of the sample, not just the QD layer, so it can affect the charge carriers as they migrate to the QD being probed. Intra-band excitation of low-energy charge carriers by the resonant laser would reduce the transport of charge to the QD region. That would affect both normal diffusion of carriers and migration via hopping between transport sites. There will be consequences for the shape parameters of the two transport processes.
\subsubsection{Super-Diffusion-Like Process}
\begin{figure}[t]
	\includegraphics[width=3.4in]{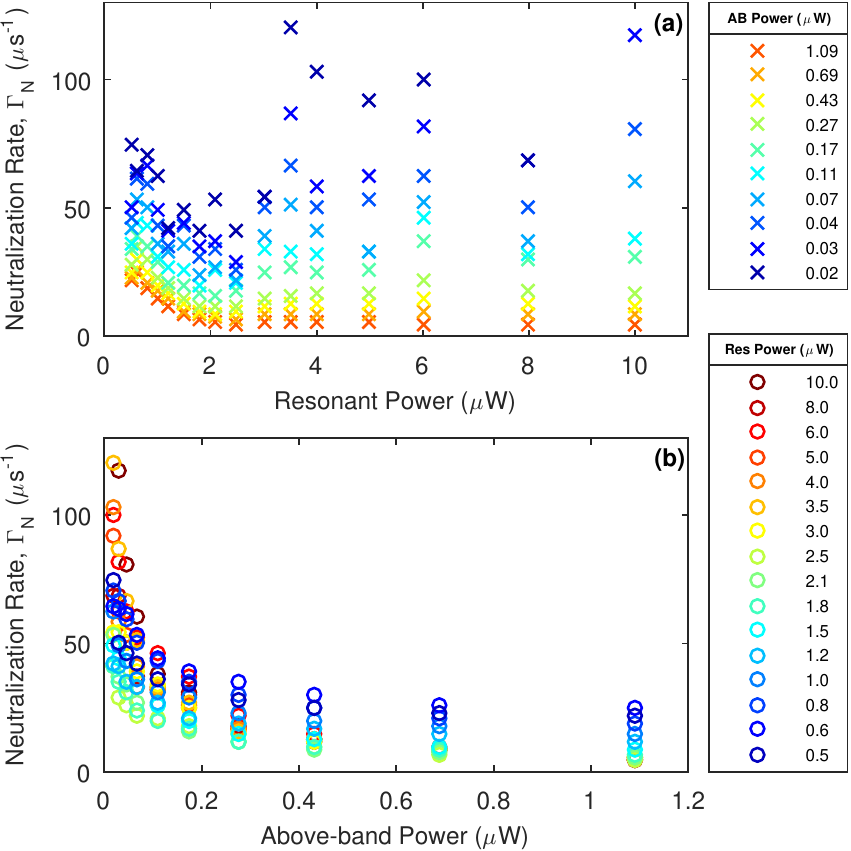}
	\caption{\label{fig:Gamma_N}  Neutralization rate per unit charge in the dot while the above-band laser is on as a function of (a) resonant laser power and (b) above-band laser power. The legends and color specify the power of the excitation laser not represented by the horizontal axis.}
\end{figure}

For the super-diffusion-like process, with increasing above-band power the growth of the capture rate is faster ($\alpha_3$ increases), the time dependence can become less compressed for lower resonant powers ($\beta_3$ decreases), and the steady-state capture rate generally increases ($C_3$ increases). Figure~\ref{fig:superdiffusion}(a-c) illustrates how the parameters of the super-diffusion-like process depend on the above-band laser. These dependencies are consistent with the expected behavior of a charge reservoir near the QD that is supplied by a combination of normal diffusion and ballistic transport of charges excited far away. A higher above-band power means more charges are excited per unit time, which will increase the rate at which the reservoir near the QD is filled, thus increasing $\alpha_3$. More charges created means that more charges will travel to the QD layer via normal diffusion as well, which will make the time dependence less compressed, thus decreasing $\beta_3$. This only occurs for lower resonant powers because higher resonant powers re-excite the normally diffusing charges, leaving ballistic transport as the dominant process and keeping the time dependence a compressed exponential. A higher excitation rate of charges will generally result in a higher steady-state population of the charge reservoir near the QD, leading to a higher steady-state capture rate, thus increasing $C_3$.

With increasing resonant power, the super-diffusion capture rate increases slower ($\alpha_3$ decreases), the time dependence becomes more compressed ($\beta_3$ increases), and the steady-state capture rate either stays about the same or increases ($C_3$ increases for high above-band powers). Figure~\ref{fig:superdiffusion}(d-f) illustrates how the parameters of the super-diffusion-like process depend on the resonant laser. The resonant laser will re-excite charges that are either in the charge reservoir near the QD or traveling there relatively slowly via normal diffusion. That will slow down the rate at which the reservoir is filled, decreasing the value of $\alpha_3$. Re-excitation will affect normally diffusing charges more strongly than ballistically moving charges because of the difference in transit time to the QD region. This will enhance the dominance of ballistic transport, making the super-diffusion-like nature of the process stronger, and increasing $\beta_3$. The steady-state capture rate has an unexpected dependence on the resonant laser power. For low to medium above-band power, the value of $C_3$ is not strongly affected by the resonant laser power. But for high above-band power, $C_3$ increases with resonant power. Perhaps for high above-band power, increasing the resonant laser power changes the ratio of electrons and holes that reach the QD region, resulting in a balance that increases the charge capture rate.
\subsubsection{Sub-Diffusion-Like Process}
For the sub-diffusion-like process, as for the super-diffusion-like process, increasing the above-band laser power will enhance the charge capture effectiveness. With increasing above-band power the capture rate increases faster ($\alpha_4$ increases), the time dependence converges on a certain stretched character ($\beta_4$ converges to $\approx 0.75$), and the steady-state capture rate increases ($C_4$ increases). Figure~\ref{fig:subdiffusion}(a-c) illustrates how the parameters of the sub-diffusion-like process depend on the above-band power. Similar to the super-diffusion-like process, a higher above-band power excites more charges and increases the rate at which the reservoir near the QD is filled, thus $\alpha_4$ increases. A higher rate of charge excitation also means a higher steady-state population of the reservoir, leading to a higher capture rate, $C_4$. The dependence of the shape parameter $\beta_4$ is more complicated because it depends on the resonant laser power. For low resonant power and low above-band power, $\beta_4$ is actually greater than 1. As the above-band power increases, $\beta_4$ decreases to about 0.75. In contrast, for high resonant power and low above-band power, $\beta_4$ is significantly less than 1. As the above-band power increases, $\beta_4$ increases to about 0.75. At high above-band power, the shape parameter is approximately 0.75 regardless of the resonant laser power.

With increasing resonant power, the growth of the sub-diffusion capture rate is slower ($\alpha_4$ decreases), the time dependence becomes more stretched ($\beta_4$ decreases), and the steady-state capture rate decreases ($C_4$ decreases). Figure~\ref{fig:subdiffusion}(d-f) illustrates how the parameters of the sub-diffusion-like process depend on the resonant power. The resonant laser has a similar effect on the sub-diffusion-like process as it does on the super-diffusion-like process. Re-excitation of charges slowly migrating through transport sites to the QD region will decrease the rate at which the reservoir of nearby transport sites are filled, thus reducing $\alpha_4$. The re-excitation will also effectively slow the hopping transport process, enhancing the stretched exponential character of the time dependence, which translates to decreasing $\beta_4$. Finally, higher resonant power means that fewer charges make it to the QD region, which will reduce the equilibrium charge population in that region and thus reduce the steady-state capture rate, $C_4$.
\subsubsection{Neutralization Process}
When the above-band laser is off, Auger recombination of the resonantly excited trion is the dominant process by which the QD becomes neutralized. In contrast, when the above-band laser is on---as during the fluorescence rise---Auger recombination is not the dominant neutralization process. The neutralization rate $\Gamma_N$ does not exhibit the dependence on resonant power that one would expect if it were solely dominated by Auger recombination processes. The dependence of $\Gamma_N$ on resonant excitation power is depicted in Fig.~\ref{fig:Gamma_N}. Recall, $\alpha_2$ in the fall model saturates with resonant power as would the time-averaged trion population. That is evidence that $\alpha_2$ is associated with neutralization of the QD via Auger recombination. However, $\Gamma_N$ \textit{decreases} with resonant excitation power for powers below 2.5 $\mu$W. Additionally, $\Gamma_N$ increases slightly at higher resonant powers for the lowest above-band powers. Thus, when the above-band laser is on, neutralization is not caused solely by Auger recombination. Neutralization may be caused by capture of an oppositely charged carrier. Additionally, carriers populating the transport sites discussed above may be excited to higher-energy continuum states via the resonant laser. The above-band excitation acts against the resonant laser by tending to charge the dot. This may explain why we see an increase in $\Gamma_N$ at high resonant powers for the lowest above-band powers.
\section{conclusion}
We studied time-resolved resonance fluorescence from a resonantly excited charged transition in an InAs QD while modulating an additional low-power above-band excitation laser. From the power dependence of the fall transients, we conclude Auger processes dominate the neutralization after the above-band excitation is turned off.
After the QD is neutralized by Auger recombination, it gets recharged from a reservoir of carriers in the local environment whose population is itself slowly decaying after the above-band excitation is turned off.
Due to the stretched exponential behavior of the associated term, we conclude this reservoir is composed of a stochastic environment of weakly attractive shallow defects acting as transport sites that charges can migrate through to ultimately be captured by the QD. 

The time and power dependence of the rise model indicates the presence of two reservoirs that supply the quantum dot with charge: one that fills with super-diffusion-like charge dynamics, and one that fills with sub-diffusion-like charge dynamics. The reservoir that fills with sub-diffusion-like charge dynamics is the same ensemble of shallow defects described in the fall section of the data. The reservoir that fills with super-diffusion-like charge dynamics is supplied by carriers created near the sample surface that ballistically travel to the neighborhood of the QD. Both reservoirs are supplied with charge via the above-band laser. The resonant laser populates the trion, causing fluorescence, but also ejects charge from the dot via Auger recombination and retards the establishment of charge in both reservoirs.

The charging and neutralization processes revealed by our analysis have implications for the design of future single photon sources. If independent electrical control of the QD charge state is impossible (as with certain optical cavities), then an above-band laser can help maintain the charge state. To have a fast time response, the QD should be relatively close to the surface where the above-band excitation is incident. A possible alternative might be to excite in the wetting layer absorption band, which would place the excited charges in the immediate locale of the QD. Though generally thought to introduce complications like spectral diffusion, the presence of defects and impurities is important to the transport of charges over long distances within the GaAs sample. Increasing the defect density by light doping or multiple layers of delta-doping might enhance the transport capability of the semiconductor host.
%
\begin{acknowledgments}
This work was supported by the National Science Foundation (DMR-1452840).
\end{acknowledgments}

\bibliography{./Library5}
\end{document}



\title{Supplemental Material: Charging dynamics of single InAs quantum dots under both resonant and above-band excitation}


\author{Gary R. Lander}
\email[]{glander@mix.wvu.edu}
\affiliation{Department of Physics and Astronomy, West Virginia University, Morgantown, WV 26506, USA}

\author{Samantha D. Isaac}
\affiliation{Department of Physics and Astronomy, West Virginia University, Morgantown, WV 26506, USA}

\author{Disheng Chen}
\affiliation{Department of Physics and Astronomy, West Virginia University, Morgantown, WV 26506, USA}

\author{Samet Demircan}
\affiliation{Department of Physics and Astronomy, West Virginia University, Morgantown, WV 26506, USA}


\author{Glenn S. Solomon}
\affiliation{Joint Quantum Institute, National Institute of Standards and Technology, \& University of Maryland, Gaithersburg, MD, USA.}

\author{Edward B. Flagg}
\email[]{edward.flagg@mail.wvu.edu}
\affiliation{Department of Physics and Astronomy, West Virginia University, Morgantown, WV 26506, USA}


\date{\today}

\pacs{need to find these}

\maketitle


\section{Anomalous Diffusion}
Throughout our analysis we utilize the shaped exponential function, $e^{-(\alpha t)^{\beta}}$, which fits well a large variety of relaxation phenomena in complex condensed-matter systems \cite{williams_non-symmetrical_1970, bohmer_nonexponential_1993, phillips_stretched_1996, klafter_relationship_1986, anderssen_kohlrausch_2004, bodunov_origin_2017, doi:10.1002/9783527622979.ch11}. When $\beta=1$, the shaped exponential reduces to the regular exponential, when $0<\beta<1$ the shaped exponential is referred to as a stretched exponential, and when $1<\beta<2$ it is referred to as a compressed exponential. Charge relaxation described with regular exponential time dependence is attributed to a regular-diffusion-like process. In 2003, Sturman \textit{et al.} described how stretched exponential relaxation in the solid state can result from charge carriers migrating through an environment of stochastically distributed transport sites, resulting in sub-diffusion-like, or slower than regular diffusion, charge relaxation \cite{sturman_origin_2003}. The compressed exponential has been less widely used to describe charge relaxation in the solid state. Charge relaxation described with compressed exponential time dependence is attributed to a super-diffusion-like, or faster than regular diffusion process. Bouchaud \cite{doi:10.1002/9783527622979.ch11} gives examples of instances where compressed exponential relaxation in the sold state is observed. The common theme is the presence of an external driving force resulting in faster than exponential charge dynamics. In the following paragraphs we will discuss why we suggest both stretched and compressed exponential relaxation exists in our sample.

\begin{figure}[t]
	\includegraphics[width=6.4in]{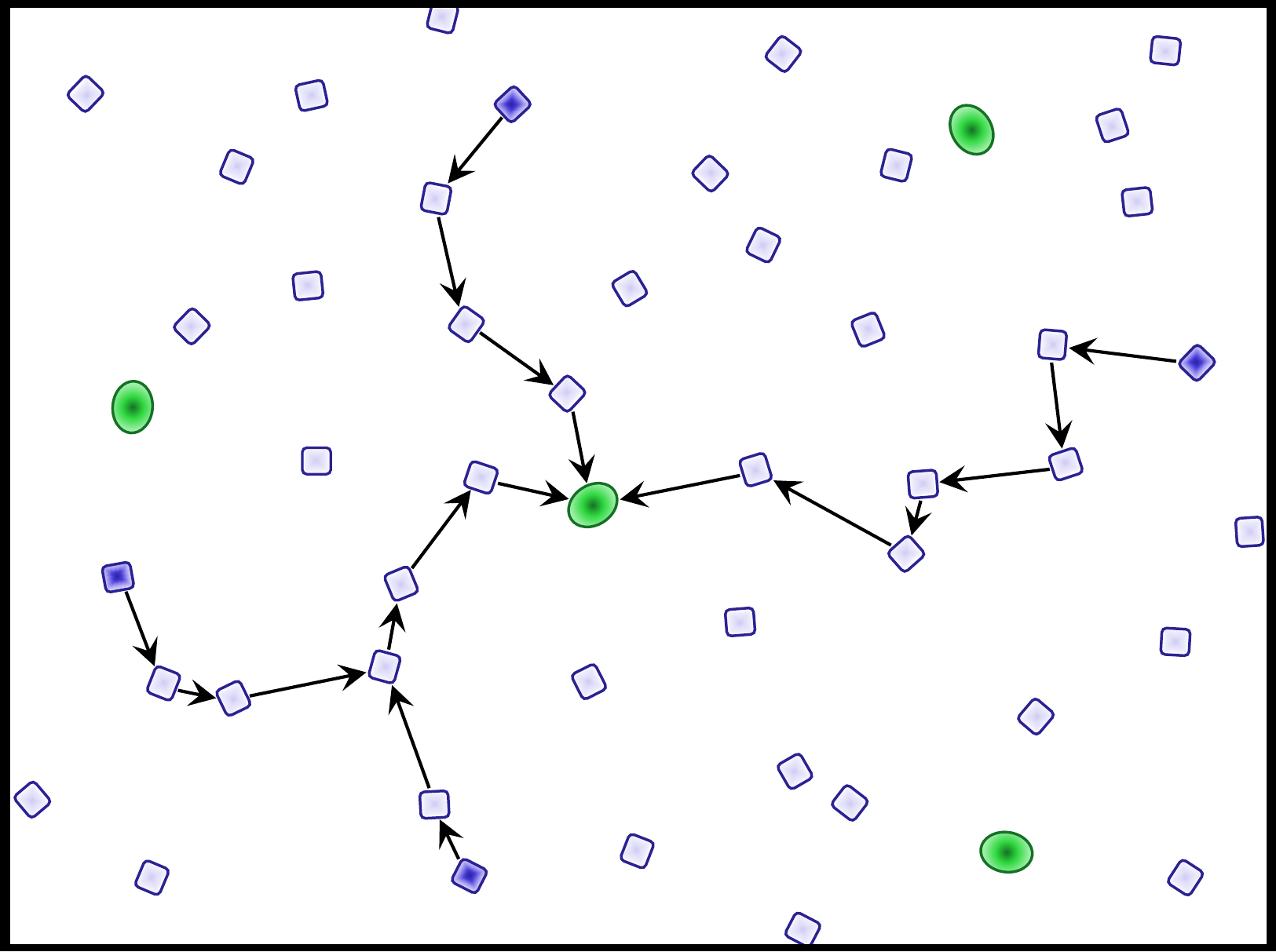}
	\caption{\label{fig:stochastic}  Schematic representation of the stochastic environment of charge transport sites and traps. The white background represents the bulk GaAs host semiconductor surrounding the QDs. Green ovals represent traps, which in our sample are InAs QDs. Blue squares represent transport sites, which are likely impurities and other shallow defects in the bulk GaAs that can weakly bind charge carriers. Dark blue squares indicate the occupation of arbitrary transport sites and the black arrows represent the charges' migration through multiple transport sites to ultimately relax into the QD being studied.}
\end{figure}

Consider an environment of randomly located deep potential wells in which charges can become electrically bound (traps) as well as a larger number of randomly located shallow wells that weakly bind charge carriers (transport sites). Charges are then introduced to the environment via some external excitation source. Carriers that weakly bind to transport sites have a probability to ‘hop’ to nearby transport sites that is proportional to the overlap of the wavefunctions corresponding to a bound charge in either trap. Ultimately, a charge may hop to a transport site adjacent to a trap and become strongly bound to the trap. The probability to make a single hop to an adjacent transport site has exponential time dependence. A stretched exponential can mathematically be represented as a sum of normal exponentials with a given weighting function \cite{johnston_stretched_2006}. The time-dependence for a charge to make multiple jumps to adjacent transport sites is a linear sum of the time-dependencies to take individual jumps, and thus the net process for charges to migrate through multiple transport sites to ultimately relax into the dot takes on stretched exponential time dependence. In our sample, migration of charge through a stochastic environment of weakly binding transport sites results in sub-diffusion-like relaxation into the QD \cite{sturman_origin_2003}. The stochastic environment of our sample is depicted in Fig.~\ref{fig:stochastic}. 

\begin{figure}[t]
	\includegraphics[width=6.4in]{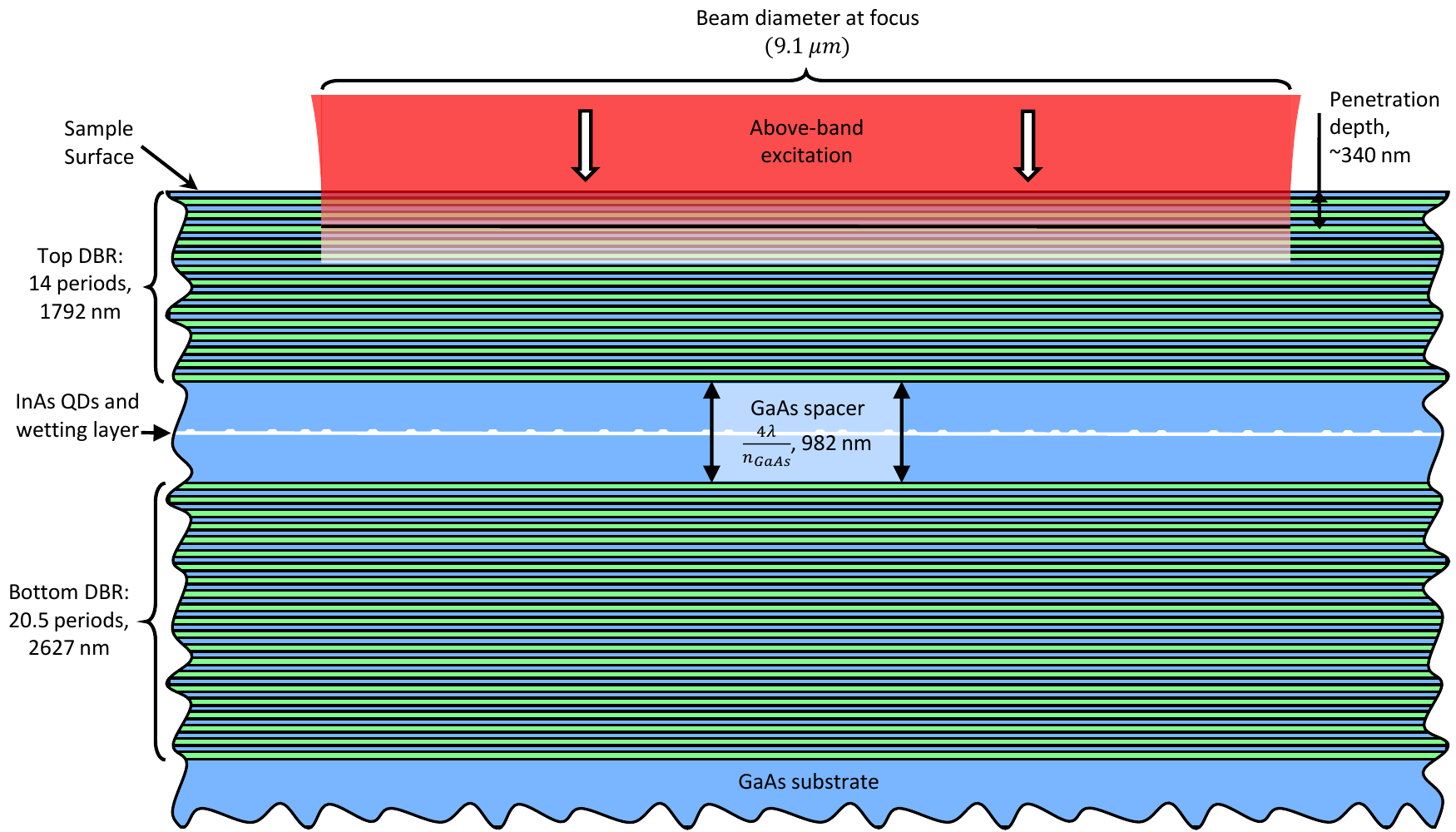}
	\caption{\label{fig:pendepth}  Schematic illustrating the skin depth of above-band photons into the sample. Red signifies the above-band laser, green AlGaAs, blue GaAs, and white InAs. The penetration depth (depth at which the intensity of the laser drops to $1/e$ its surface value) is depicted by the black line within the focused laser spot. The laser illustration is extended to the skin depth (depth at which the laser intensity drops to $1/e^2$ of it's surface value. All components are to scale except the InAs QDs and wetting layer.}
\end{figure}

The timescale for the rise transient of the time-resolved resonance fluorescence indicates the charges excited by the above-band laser are created nonlocal to the QD layer. At the sample surface sits a distributed Bragg reflector consisting of alternating layers of AlGaAs and GaAs. Propagating through this heterostructure above-band photons (632 nm) have a penetration depth (depth at which the intensity of the laser drops to $1/e$ of its surface value) of approximately 340 nm \cite{rakic_modeling_1996}. However, the QD layer sits approximately 2300 nm below the sample surface. The carriers created via absorption of above-band photons move with ballistic-like motion due to their high kinetic energies. This results in the carriers spreading throughout the sample with super-diffusion-like characteristics while they are hot and regular-diffusion-like characteristics once they cool \cite{doi:10.1002/9783527622979.ch11, morishita_compressed_2012}. Some carriers have trajectories that set them near the QD layer when they cool to the gamma point of the GaAs host and can become bound to QDs or the weakly-binding transport sites described above. A to-scale schematic of the sample is shown in Fig.~\ref{fig:pendepth}.
\section{Additional Parameter Dependencies}
During the fall section of the data the above-band excitation is turned off and the fluorescence drops to zero as the dot neutralizes. Note in the following figures the above-band powers specified are those before the above-band excitation is turned off. The time-resolved resonance fluorescence of the fall section was fit with the following phenomenological function:
\begin{equation}
I=A_1e^{-\alpha_1t}+A_2e^{-(\alpha_2t)^{\beta_2}} \label{eqn:fall} \\                                                               
\end{equation}
\begin{figure}[t]
	\includegraphics[width=3.4in]{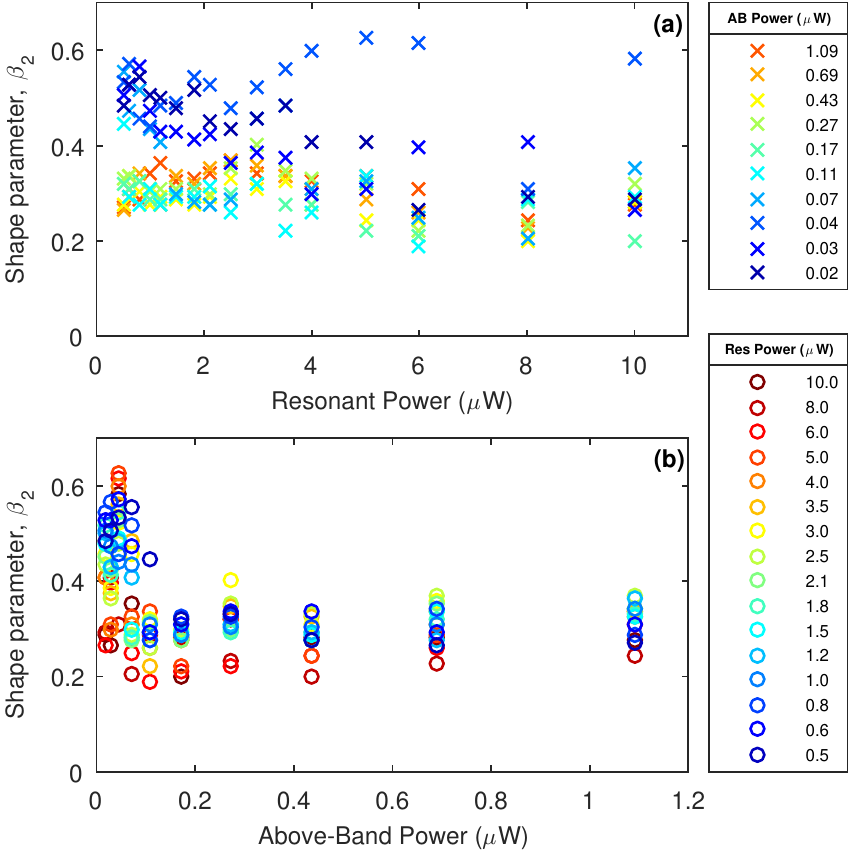}
	\caption{\label{fig:beta2}  The shape parameter $\beta_2$ as a function of (a) resonant and (b) above-band laser power. Note $\beta_2$ always relaxed to a value between 0 and 1 indicating association with a sub-diffusion-like process. The legends and color specify the laser power not depicted by the horizontal axis.}
\end{figure}

Where $A_1$ and $A_2$ are amplitudes with $A_1+A_2$ equal to the steady-state fluorescence intensity before the above-band excitation is turned off, $\alpha_1$ is a neutralization rate while $\alpha_2$ is a scale parameter, and $\beta_2$ a shape parameter. The first term in the above equation is attributed to ejection of charge from the QD via Auger recombination. $\beta_2$ always obtained a value between 0 and 1 during the fitting, indicating sub-diffusion-like depletion of the associated charge reservoir. After the above-band excitation is turned off, the reservoir consisting of the stochastic distribution of transport sites does not deplete instantly and charges from the reservoir can still be captured by the QD while a charge population in the reservoir exists. The second term fits the long-time, slowly-sloping tail in the resonance fluorescence intensity. We attribute the second term in the above equation to \textit{recharging} of the QD from the reservoir before it is depleted of charge. Thus, the second term in the above equation represents the charge population of the reservoir that is comprised of the stochastic distribution of transport sites after the above-band excitation is turned off.   
\begin{figure}[t]
	\includegraphics[width=6.4in]{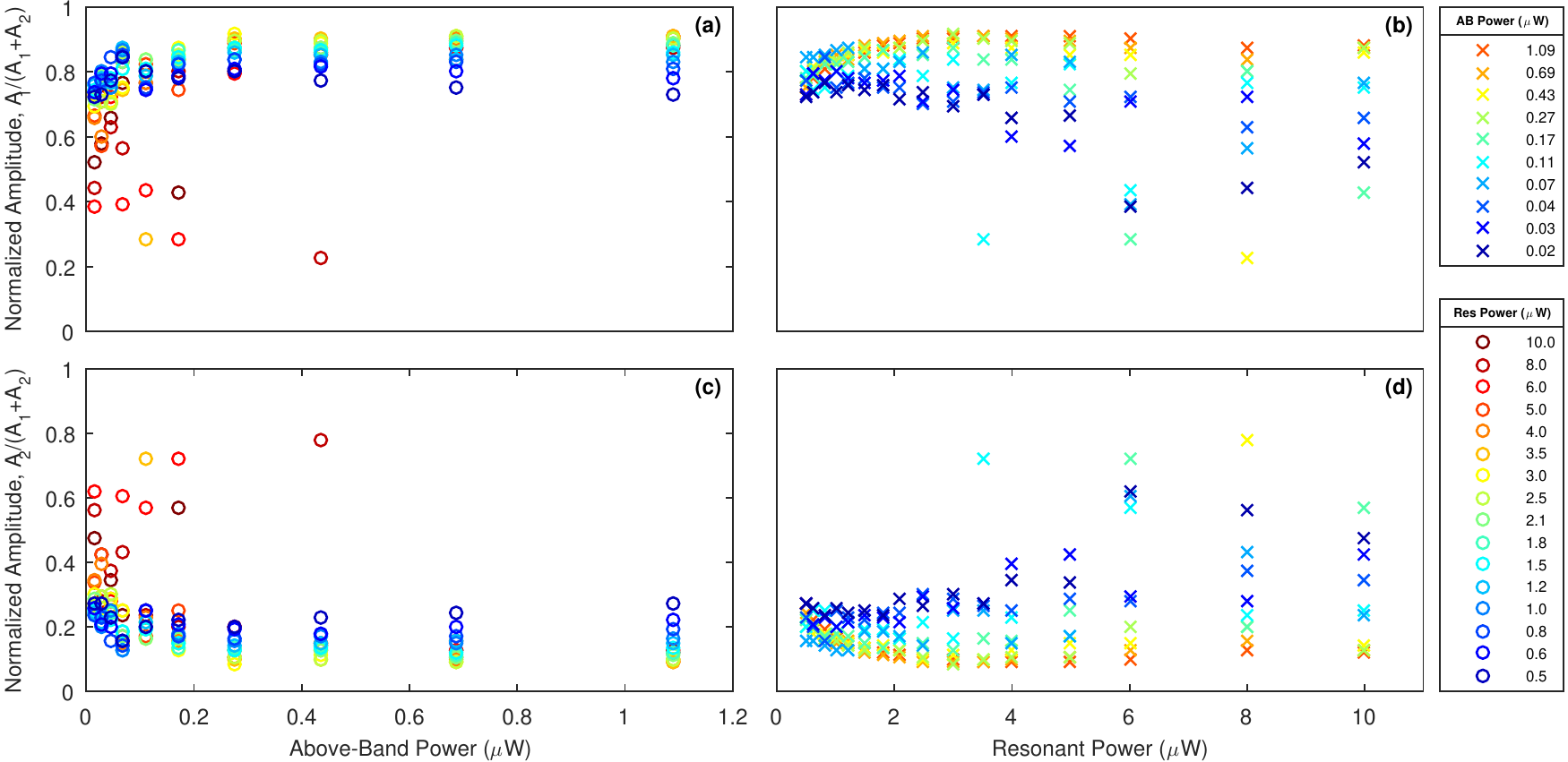}
	\caption{\label{fig:FallAmps}  The normalized amplitude associated with Auger recombination, $A_1$, is depicted as a function of (a) above-band and (b) resonant excitation power. The normalized amplitude associated with the slow depletion of the sub-diffusion reservoir, $A_2$, depicted as a function of (c) above-band and (d) resonant excitation power. The legends and colors specify the power of the other excitation laser not depicted by the horizontal axis.}
\end{figure}

The shape parameter, $\beta_2$, is depicted in Fig.~\ref{fig:beta2}. Although $0<\beta_2<1$, two behavioral regimes exist for $\beta_2$ as a function of above-band laser power. For powers below 0.07 $\mu$W, the shape parameter is about 0.5. As the above-band power increases past that threshold, $\beta_2$ rapidly switches to values around 0.3. 

Finally, the amplitudes $A_1$ and $A_2$ are depicted in Fig.~\ref{fig:FallAmps}. $A_1\approx4A_2$ throughout most of acquisition space, except for the lowest above-band powers. This indicates neutralization of the QD via Auger recombination happens at roughly four times the rate at which charges deplete from the reservoir through the QD after the above-band excitation is turned off.

\bibliography{./Library5}